\documentclass[aps,prx,twocolumn,superscriptaddress,nofootinbib,nobalancelastpage]{revtex4-2}
\usepackage{graphics}
\usepackage{amsmath}
\usepackage{physics}
\usepackage{graphicx}
\usepackage{color}
\usepackage{amsfonts}
\usepackage{amssymb}
\usepackage{float}
\usepackage{hyperref} 
\usepackage{tabularx}

\usepackage{xcolor}

\usepackage{tabularray}
\usepackage{colortbl,multirow,array}

\usepackage{flushend}

\DeclareMathOperator*{\argmin}{arg\,min}
\DeclareMathOperator*{\argmax}{arg\,max}

\setcitestyle{authoryear,round,semicolon} 
\usepackage{titlesec}

\titleformat*{\section}{\normalsize\bfseries\sffamily}
\titleformat*{\subsection}{\normalsize\bfseries\sffamily}
\titleformat*{\subsubsection}{\normalsize\bfseries\sffamily}

\makeatletter
\renewcommand{\p@subsection}{}
\renewcommand{\p@subsubsection}{}
\makeatother

\begin{document}

\title{Quantum Adversarial Learning for Kernel Methods} 

\author{Giuseppe Montalbano} \email{giu.montalbano@gmail.com}
\affiliation{
Ca’ Foscari Challange School, University Ca’ Foscari, Dossoduro 3246,
Venezia, 30123, Italy.
}

\author{Leonardo Banchi}\email{leonardo.banchi@unifi.it}
\affiliation{Department of Physics and Astronomy, University of Florence,
via G. Sansone 1, I-50019 Sesto Fiorentino (FI), Italy}
\affiliation{ INFN Sezione di Firenze, via G. Sansone 1, I-50019, Sesto Fiorentino (FI), Italy }

\date{\today}

\begin{abstract}
	We show that hybrid quantum classifiers based on quantum kernel methods and
	support vector machines are vulnerable against adversarial attacks, 
	namely small engineered perturbations of the input data can deceive the classifier into 
	predicting the wrong result. Nonetheless, we also show that simple defence strategies based on data augmentation with 
	a few crafted perturbations can make the classifier robust against new attacks. 
	Our results find applications in security-critical learning problems and in mitigating 
	the effect of some forms of quantum noise, since the attacker can also be understood as part of the
	surrounding environment. 
\end{abstract}

\maketitle

\section{Introduction  }\label{sec:intro}

With the advance of artificial intelligence and its adoption in strategic, safety-critical and 
healthcare applications, it has become essential to study the
possible vulnerabilities that affect the different types of algorithms, and to
develop appropriate countermeasures. Because of this, the possibility of deceiving machine
learning algorithms by means of intentionally crafted data as well as the
design of proper defense strategies has been investigated quite actively in
classical machine learning, giving rise to the research field of adversarial
machine learning \citep{biggio2018wild}. 

Adversarial machine learning studies many ways of attacking and defending
machine learning algorithms with different kinds of attacks, that can be
characterized based on three aspects: i)
the \emph{time} when the attack is performed, and whether the algorithm or the model is attacked;
ii) the \emph{amount of information} needed to perform the attack, e.g.~in
so called white-box attacks we have complete details about the algorithm and
the data, in contrast to black-box attacks where the algorithm and the model is unknown to the attacker; 
iii) the \emph{goals of the attack}, e.g.~an untargeted attack, when 
the attacker wants to cause just a wrong behaviour, or a targeted 
attack, when they want to achieve a specific result.
Input data that are purposely crafted by an attacker to induce the
model to make mistakes are generally called adversarial examples or adversarial samples. It
has been shown that, in some cases, these adversarial examples can take the
form of imperceptibly small perturbations to real input data, such as making a
human-invisible change to pixels of an image \citep{szegedy2014intriguing}.

Given the remarkable progresses in quantum computing hardware, different machine learning algorithms 
and techniques have been adapted to exploit the computation capabilities offered 
by these systems, giving rise to the field of Quantum Machine Learning (QML). However, due to the limitations of current 
devices, such an adaptation often requires a significant algorithmic redesign to streamline
the number of operations done in quantum device itself. 
Therefore, it is not always trivial to adapt tricks and results from the classical literature 
to the quantum case. 
Two popular quantum machine learning algorithms are based on Quantum Neural Networks (QNN)
and Quantum Kernel Methods (QKM), see e.g.~\citep{schuld2015introduction}. 
The vulnerability of quantum neural networks against adversarial attacks and the development of 
defence strategies have been investigated both theoretically
\citep{lu2020quantum,liu2020vulnerability} and experimentally \citep{ren2022experimental}.

In this work we show the vulnerability of quantum kernel methods against similar attacks, 
and develop a simple defence strategies based on data augmentation with the crafted adversarial samples. 
We focus on 
binary classification tasks for classical data, though similar considerations can
be done for the classification of quantum data, which 
has different applications \citep{gebhart2023learning}. In the latter case,
perturbations needs to be directly applied to the quantum state. 
Our results are supported by both analytical arguments and extensive numerical simulations,
performed using the Qiskit library \citep{Qiskit}.

Our results find application in safety-critical learning problems, 
but also for the development of more robust learning techniques against quantum 
noise. Indeed, a subset of the unavoidable errors in current quantum devices can be 
understood as a perturbation in the inputs. While we cannot fix general hardware errors 
without sophisticated error correction or mitigation techniques, we can make 
the model more robust against input perturbations, and, as a result, against 
fluctuations in the parameters of the encoding gates. 
To show that by developing robust algorithms 
against adversarial inputs, we can also mitigate the effect of some hardware noise,
we have performed a proof-of-principle experiment on a real device, 
using the IBM Quantum Platform.

\section{Background}
In this section we provide a mathematical introduction to 
adversarial learning and kernel methods. 

\subsection{Quantum adversarial machine learning}\label{sec:adversarial}


Let $X\subset \mathbb{R}^d$ and $Y$ denote, respectively, the space of classical inputs and discrete labels. 
In supervised learning tasks on classical data, we have access to a training dataset
$D:=\{\mathrm{x}_{i},\mathrm{y}_{i}{\}}_{i=1}^{M}$ with $M$ pairs, where $\mathrm{x}_{i}\in X$
are input vectors and $\mathrm{y}_{i}\in Y$ are the associated labels. The
input vectors and the labels are related by some unknown function $f: X\mapsto
Y$. The goal of the supervised learning task we consider is to train a model
$h\left(\cdot,\theta \right): X\mapsto Y$ to approximate $f$, where
the vector $\theta$ defines the model parameters, belonging to some appropriate
subset of $\mathbb{R}^{N}$.

To train the model we define a loss function 
$\mathcal{L}\left(h\left(\cdot,\theta \right),\cdot\right){:}~(X{\times}Y)\mapsto\mathbb{R}$ 
that introduces a penalty whenever the predicted label is different from the true one,
i.e. $h(x,\theta)\neq f(x)$. We then define the empirical loss 
\begin{equation}
	\mathcal L(D,\theta) = \frac 1M \sum_{i=1}^M \mathcal{L}\left(h\left(\mathrm{x}_{i},\theta \right),y_{i}\right),
	\label{eq: empirical loss}
\end{equation}
which measures, on average, how well the model approximates $f$ on $D$. 
The training problem can be
formulated as a minimization problem:
\begin{equation}
	\theta _{\rm opt}= \argmin_{\theta} \mathcal L(D,\theta). 
	\label{eq: training}
\end{equation}
In kernel methods the above optimization problem
is convex and therefore one is guaranteed that the optimal 
parameters $\theta_{\rm opt}$ define a global optimum. This is a
strong theoretical advantage of kernel-based QML algorithms compared to QNNs.
Once the training process is completed, namely the above  optimization is
solved, the model is evaluated on a test dataset of unseen input vectors to
check its generalization capabilities.

The type of attack that we consider in this study belongs to the category of
white-box untargeted attacks, applied against the model of a quantum classifier. In
this scenario, the technique to generate an adversarial sample starting from a true 
sample $x_i$ consists on
leveraging the same loss function used during the training of the model, 
though in
this case the parameters $\theta $ are kept fixed at their optimal value
$\theta_{\rm opt} $ 
found during training, and we maximize over the input space:
\begin{equation}
\delta _{i,\rm adv}=\argmax_{\delta \in \Delta }\mathcal{L}\left(h\left(\mathrm{x}_{i}+\delta ,\theta _{\rm opt}\right),y_{i}\right),
\label{eq: adversarial}
\end{equation}
where $\Delta $ represents a limited region of the input space where to look for a solution, in order to keep the perturbation small.

\subsection{Quantum-enhanced support vector machines}

Machine learning techniques based on kernel methods are a family of techniques
that builds on the idea of using a particular measure to evaluate the
similarity between vectors in some appropriate Hilbert space with the objective
to solve some task \citep{schuld2021machine}. In this work we focus on binary classification via 
Support Vector Machines (SVM)
\citep{cristianini2000introduction}
where training consists in finding a decision hypersurface on a Hilbert space, called \emph{feature space}, 
and the classification is done by checking on which side of the plane the input belongs. 
QML algorithms based on kernel methods and SVM have gained significant attention as a
potential candidate for achieving a possible quantum advantage over classic
algorithms \citep{liu2021rigorous}. 

For simplicity we focus on binary classification problems with $Y=\{+1,-1\}$.
A nonlinear SVM classifier makes decisions about the class of an input sample
using an optimal hypersurface in the input space, which is mapped into a hyperplane in the feature space. The latter is defined as
\begin{equation}
\langle w,\phi \left(x\right)\rangle +b=0,
\label{eq:hypersurface}
\end{equation}
where $\phi(x)$ is the \emph{embedding} function that maps the input to the feature space. 
The optimal hyperplane is selected by solving a quadratic programming, a convex optimization
problem with inequality constraints, described in its dual problem formulation
as follows:
\begin{equation}
\max_{\alpha }\left[\sum_{i=1}^{M}\alpha _{i}-\frac{1}{2}\sum _{i,j=1}^{M} 
y_{i}y_{j}\alpha _{i}\alpha _{j}k(x_i,x_j)\right],
\label{eq: dual opt}
\end{equation}
with constraints ${\sum }_{i=1}^{M}\alpha _{i}y_{i}=0$            and
$0\leq \alpha _{i}\leq C$ for $i=1\ldots M$, where
$C$ is an hyper-parameter used as a penalty weight to
address samples in the training dataset that would not allow a linear
separation in the considered feature space. The kernel 
\begin{equation}
	k(x_i,x_j) = \langle \phi \left(x_{i}\right),\phi \left(x_{j}\right)\rangle,
	\label{eq: kernel}
\end{equation}
is the (euclidean) inner product between two feature vectors, obtained via an
embedding function $\phi$.
The convex optimization problem in Eq.~\eqref{eq: dual opt} can be solved with
open-source libraries \citep{chang2011libsvm} with polynomial complexity in $M$. 
From the solution of the dual problem, it is possible to show that 
$ w=\sum _{i=1}^{M}y_{i}\alpha _{i}\phi \left(x_{i}\right)$, with a related 
expression for $b$ (see below). 
Once the optimal hypersurface has been found, 
the nonlinear SVM classifier predicts the label of an input sample based on the sign of the decision function $f\left(\mathrm{x}\right)$:
\begin{align}
\label{eq: f}
	f(x)&=\left\langle w,\phi \left(x\right)\right\rangle +b = 
		\\&=\sum _{i=1}^{M}\left[\alpha _{i}y_{i}k\left(x_{i},x\right)+ \frac{y_i -
			\sum_{j=1}^M\alpha_jy_jk(x_i,x_j)}M\right].\nonumber
\end{align}
When $f\left(x\right)\geq 0$ the classifier predicts a label $+1$, otherwise it predicts a label $-1$.
We observe that in the dual formulation, both the training and test processes,
Eqs.~\eqref{eq: training} and \eqref{eq: f}, do not require any evaluation in the feature space 
with the exception of the kernel \eqref{eq: kernel}, while the direct optimization of $w$ in the primal 
formulation might be complex for very large feature spaces.

Quantum circuits can be exploited to define a feature space with a 
dimension that increases exponentially with the number of qubits $n$, which may 
allow an easier separation of inputs with different labels. Let
$U\left({x}\right)$ be a unitary circuit applied to an
initial state $| 0^{n}\rangle $ to produce $\left| \phi
\left({x}\right)\rangle =U\left({x}\right)\right| 0^{n}\rangle $. 
Following \citep{schuld2021machine} we define the encoding feature map $\phi:\mathbb{R}^d\mapsto\mathcal F$
as the function mapping 
each input sample ${x}\in \mathbb{R}^{d}$ to the density matrix
	$\phi \left({x}\right)=\left| \phi \left({x}\right)\rangle \langle \phi \left({x}\right)\right| 
	=:\rho \left({x}\right)$,
where $\mathcal{F}$ is the space of density matrices of size $2^{n}\times
2^{n}$, equipped with the Hilbert-Schmidt inner product $\langle \rho ,\sigma
\rangle _{\mathcal{F}}=\Tr\left(\rho ^{\dagger }\sigma \right)$ for $\rho
,\sigma \in \mathcal{F}$. 
With these definitions the kernel \eqref{eq: kernel} 
\begin{equation*}
k\left({x}_{1},{x}_{2}\right)=\Tr\left(\rho \left({x}_{2}\right)\rho \left({x}_{1}\right)\right)=\left| \langle \phi \left({x}_{2}\right)\right| \phi \left({x}_{1}\right)\rangle | ^{2},
\end{equation*}
is nothing but the fidelity between two quantum state embeddings. Although such embeddings map the inputs onto 
a space that grows exponentially with the number of qubits, by exploiting the kernel trick, 
we only need to know how to evaluate the kernel
values between feature vectors corresponding to samples mapped from the input
dataset. This allows us to avoid optimization 
problems in a huge Hilbert space, and to work directly 
with the space of the training dataset that, having a 
lower dimension, is easier to manipulate. 

A Quantum-enhanced Support Vector Machine (QSVM) is a hybrid
classical-quantum algorithm that exploits a quantum system to
compute the quantum kernel values and uses a classical SVM algorithm to
determine the optimal decision hypersurface. There are different techniques to 
estimate the quantum kernel by means on a quantum computer. Examples include a
technique commonly referred to as Quantum Kernel Estimation (QKE) \citep{havlivcek2019supervised,liu2021rigorous},
which consists in concatenating the two circuits, followed by an estimation of the probability to 
measure all qubits in the state $|0\rangle$. 
Accordingly, for a single kernel value, we have to repeat the measurement many
times in order to obtain a good estimation. 
For embedding circuits with many layers, alternative circuits 
such as the swap test might be more appropriate \citep{schuld2021machine}. 

The quantum embedding circuit that we use to encode input
data into quantum states completely describes the associated
quantum kernel and therefore characterizes the performance achievable by the
SVM classifier. A general goal is, therefore, to look for a kernel
that guarantees the best performance and that best fits the dataset at
hand. Since the optimal kernel is unknown, we use the kernel alignment technique described 
in appendix~\ref{sec:kernel alignement} to optimize the quantum circuit, and hence 
the kernel.

\section{Adversarial Learning on Quantum-Enhanced Support Vector Machines}\label{sec:qadv}  

Following the general scheme introduced in Sec.~\ref{sec:adversarial}, 
 we now introduce a technique to perturbate a
legitimate input sample to deceive a previously trained QSVM classifier. In
addition to the general concepts described previously, we have to consider a couple of basic aspects:
\begin{itemize}
\item Knowing the principles on which a SVM builds on, the goal of the
	attacker is to generate adversarial perturbations to legitimate samples, 
	so that the resulting feature vectors (density matrices) are located on
	the other side of the decision hypersurface, with respect to the 
	unperturbed ones. 
	This type of
	attack is referred to in literature as Evasion Attack \citep{li2022kernel}. 
\item The perturbations to the original samples should be as ``light'' as
	possible, so we would like to keep the distance between the original sample
	${x}$ and its modified version ${x}_{\rm adv}$ small. A possible
	way to evaluate this distance, is to use the euclidean distance
	$\| {x}-{x}_{\rm adv}\|_{2}$.
\end{itemize}

\begin{figure}[t]
	\centering
	\includegraphics[width=0.45\textwidth]{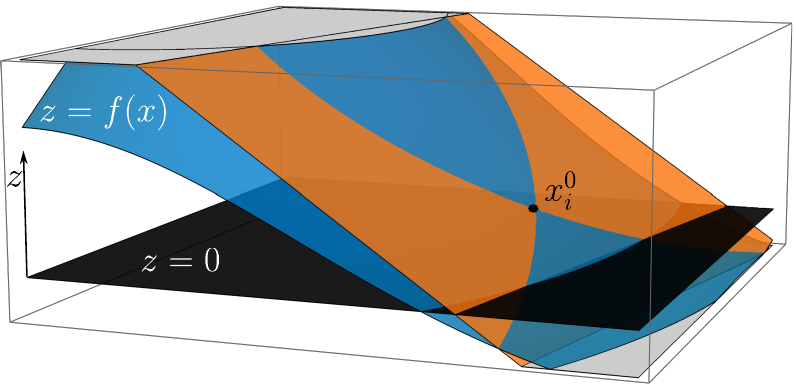}
	\includegraphics[width=0.45\textwidth]{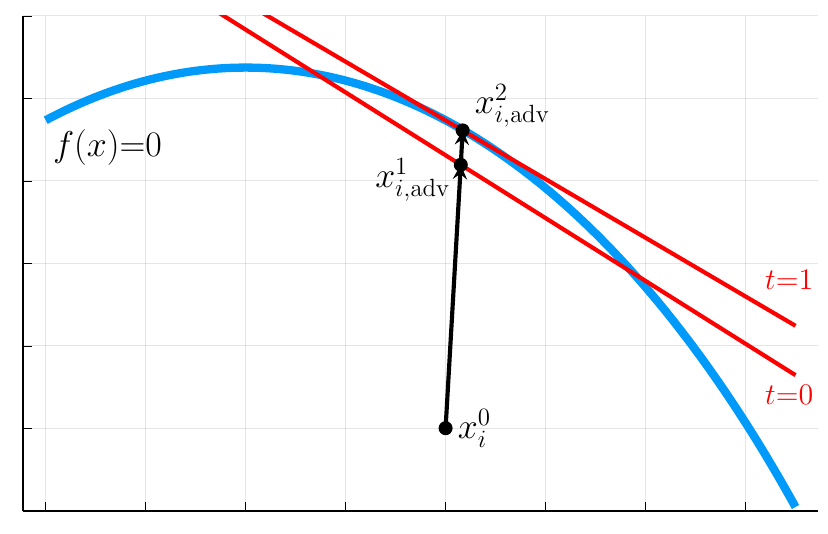}
	\caption{
		(top) Three dimensional representation of the the decision function $f$ (blue) for a two-dimensional 
		input $x$ and an extra axis $z$, together with the tangent plane near $x_i^0$ (orange). 
		(bottom) The intersection of the above surfaces with the $z=0$ plane (in black in the top figure), together with 
		two steps of the iteration described by 
		Eqs.~\eqref{eq: gradient iteration} and \eqref{eq:adaptive}. The blue line is the decision 
		hypersurface $f(x)=0$, while the orange lines are the 
		approximations of the hypersurface at iterations $t=0$ and $t=1$, using hyperplanes -- 
		that for $t=0$ is a projection of the orage hyperplane from the top figure.
	}
	\label{fig:3dpict}
\end{figure}

To enforce these two constraints simultaneously, we may focus on Eq.~\eqref{eq: adversarial} and 
add a penalty term $\|x-x_{\rm adv}\|^2$, in place of the hard constraint $\delta\in\Delta$. 
Assuming that $x_{\rm adv,i} = x_i + \delta_i$ with a small $\delta_i$, we find 
\begin{align}
	\delta _{i,\rm adv}&=\argmax_{\delta_i  }\mathcal{L}\left(f(\mathrm{x}_{i}+\delta_i ),y_{i}\right) 
	- \frac{\|\delta_i\|^2}{2\eta},
									\\ &\approx
\argmax_{\delta_i  }\nabla_x\mathcal{L}\left(f(\mathrm{x}_{i}),y_{i}\right) \cdot \delta_i
- \frac{\|\delta_i\|^2}{2\eta},
\end{align}
where $\eta$ is a positive parameter to tune the strength of the penalty term. 
In SVM $\mathcal L$ is the hinge loss $\mathcal L(f(x),y) = \max\{0,1-yf(x)\}$, hence 
the above problem has the following solution 
\begin{equation}
	\delta_{i,\rm adv} \approx -\eta y_i \nabla_x f(x_i),
	\label{eq: delta adv}
\end{equation}
which can be interpreted as a single gradient step with learning rate $\eta$. 
Geometrically, the value of $\nabla _{{x}}f$ is a vector normal to intersection line between the hyperplane tangent to the hypersurface defined by the decision function $f\left({x}\right)$ at ${x}_{i}$ and the hyperplane z=0, see Fig.~\ref{fig:3dpict}.
Note also that the hinge loss is analytic away of the decision hypersurface. 
Leveraging these observations, we can implement an iterative procedure that
starts from an initial point ${x}_{i}$ and shifts that point toward the
separating  hypersurface, using the gradient of the decision function as
the direction to update the point at each step. 
By repeating the above procedure, the adversarial perturbation at the $t$-th iteration 
satisfies 
\begin{equation}
	x_{i,\rm adv}^{t+1} = 
	x_{i,\rm adv}^{t} - \eta_i^t y_i \nabla_x f(x_{i,\rm adv}^t),
	\label{eq: gradient iteration}
\end{equation}
with $x_{i,\rm adv}^{0} =x_i$ and where $\eta_i^t$ is an adaptive learning rate -- 
in the simplest case $\eta_i^t\equiv \eta$.
The iteration is stopped once $\rho(x_{i,\rm adv}^t)$ 
is located on the other side of the decision hypersurface,
or the maximum number of iterations is reached. The latter must be 
kept small to guarantee a small $\|x_i-x_{i,\rm adv}\|$.

A better iterative technique can be defined by exploiting the geometrical 
structure of SVMs. To clarify this, we start with linear SVM, where $\phi(x)\equiv x$ and 
the decision hypersurface \eqref{eq:hypersurface} 
can be written as a hyperplane $f(x) ={w}^{T}{x}+b=0$. The vector ${w}$ is
normal to the hyperplane, and
$|f(x_i)|/\|w\|$ is the distance between a general point $x_i$ and the hyperplane. 
Leveraging this observation, to project a point into the plane we should add a perturbation along 
the orthogonal direction, identified by the normalized vector $w/\|w\|$,  with a strength given by the distance.
This results in 
\begin{equation}
	\delta_i = -\frac{w}{\|w\|}\frac{|f(x_i)|}{\| w\|}.
	 \label{eq:gradient linear}
\end{equation}
In the general case, where the embedding map $\phi$ is non-linear, $f(x)=0$ identifies a
hypersurface, which can be approximated as a hyperplane around a certain point $x_i$ as
$f(x_i+\delta_i)\approx f(x_i)+ \nabla_x f(x_i)^T \delta_i$, as shown in  Fig.~\ref{fig:3dpict}. 
Therefore, the vector $w$ 
from the previous analysis can be replaced with the gradient of $f$.
Combining this intuition with Eq.~\eqref{eq:gradient linear}, in the non-linear case 
the formula to find an adversarial perturbation becomes equivalent 
to the gradient step \eqref{eq: delta adv}, but with an {\it adaptive} learning rate
\citep{li2022kernel}
\begin{equation}
	\eta^t_i = \frac{|f(x^t_{i,\rm adv})|}{\|\nabla_x f(x^t_{i,\rm adv})\|^2}. 
	\label{eq:adaptive}
\end{equation}
Since the approximated decision hyperplane changes during the iteration, due to the curvature of the hypersurface, 
the above procedure must be repeated multiple times, as in \eqref{eq: gradient iteration}. 
A pictorial representation of this iteration is shown in Fig.~\ref{fig:3dpict}. 
When $x_{i,\rm adv}^t$ approaches the decision hypersurface, $\eta_i^t$ can get very small. In 
that case, we may empirically replace the adaptive learning rate \eqref{eq:adaptive} with a larger constant $\eta$.

From the above analysis it is clear that the best way to generate adversarial samples is to start 
either from inputs $x_i$ with $|f(x_i)|\approx 0$, which are close to the hypersurface and can switch side 
with few gradient steps, or from inputs with a large gradient. 
When quantum kernels are built via deep quantum circuits, then from the kernel concentration phenomenon 
\citep{thanasilp2022exponential}, we can expect that $|f(x_i)|\approx 0$ for a large number of inputs,
making the generation of adversarial samples easier.

This technique to generate adversarial samples builds on the assumption that we
can compute the gradient of $f$, which, thanks to the analytic expression Eq.~\eqref{eq: f},
requires the ability to compute the gradient of the quantum kernel. Since the kernel depends linearly 
on the state $\rho(x)$, it is possible to use techniques such as  the parameter-shift 
rule \citep{mitarai2018quantum,banchi2021measuring,schuld2021machine} to express the $\ell$th component 
of the gradient as 
\begin{equation}
	\partial_{x_\ell} f\left({x}\right)=\sum_{i=1}^{M}\alpha _{i}y_{i}\left[
k\left({x}_{i},x+\pi_\ell\right)-
k\left({x}_{i},x-\pi_\ell\right)\right],
\end{equation}
where the vector $\pi_\ell \in\mathbb{R}^d$ has components $(\pi_\ell)_i=\frac{\pi}{4} \delta_{i,\ell}$. 
When the embedding circuit contains gates that cannot be simply expressed via 
Pauli rotations, it is necessary to use 
extensions such as the stochastic parameter-shift rule \citep{banchi2021measuring}.

\section{Results}
\label{sec: results}

\begin{figure}[t]
	\centering
	\includegraphics[width=0.5\textwidth]{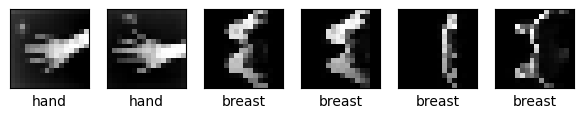}
	\caption{Some samples from the training and test dataset.}
	\label{fig:images}
\end{figure}
\begin{table}[t]
	\centering
	\begin{tabular}{|l|l|l|l|}
		\hline
		& \textbf{Training set~~} & \textbf{Test set~~} & \textbf{Total~~} \\
		\hline
		\textbf{Hand~~} & 245 & 45 &290 \\
		\hline
		\textbf{Breast~~} & 255 & 55 & 310 \\ 
		\hline
		\textbf{Total} & 500 & 100 & 600 \\
		\hline
	\end{tabular}
	\caption{Composition of the training and test datasets.}
	\label{tab:traintest}
\end{table}

In our numerical study, we use
one of the datasets exploited in the work on adversarial training with quantum neural networks
\citep{ren2022experimental}, so we can compare the results.
That dataset is publicly available
and consists of images from the Medical MNIST, a collection of standardized
biomedical images. Some examples are shown in Fig.~\ref{fig:images}. 
In the experiments we use a subset composed of 600
monochromatic images, consisting of Magnetic Resonance Imaging data of two classes: hand and breast, divided as 
in Table~\ref{tab:traintest}.
The images are scaled to the resolution of 16 by 16 pixels and then normalized.


After some preliminary experiments we decided to use a
quantum embedding circuit with a structure very similar to the one used in \citep{ren2022experimental}:
we designed a quantum circuit with the same number of qubits, 10, to efficiently map the input data
into a quantum state, using some tunable parameters and some entangling layers, 
implemented with CNOT gates.

We map the $16\times16$ pixels of each image in 10 qubits using 
three blocks of angle encoding layers:
i) 8 alternating layers of Rx and Rz gates;
ii) 12 alternating layers of Rx and Rz gates;
iii) 6 alternating layers of Rx and Rz gates.
Two layers of CNOT gates between nearest qubits are added after each block of angle encoding gates 
to create entanglement.
Before data encoding, each input sample is extended with four zero components so to map exactly into the 260 angle encoding gates.
To define a tunable quantum embedding we also introduce 30 variational parameters $\theta_i$ with $i=0\ldots 29$.
To reduce the total number of angle encoding gates in the quantum circuit, these parameters are summed up to the first 30 components of the input data $x_i$ and encoded in the first three layers of gates.
This is equivalent to consider each of the first three layers of gates as actually composed of two layers of rotation gates each, where input data and tunable parameters are encoded into dedicated angle encoding gates.
From now on, we will call such an encoding circuit the \textit{large feature map}.

\begin{figure}[t]
	\centering
  \includegraphics[width=0.48\textwidth]{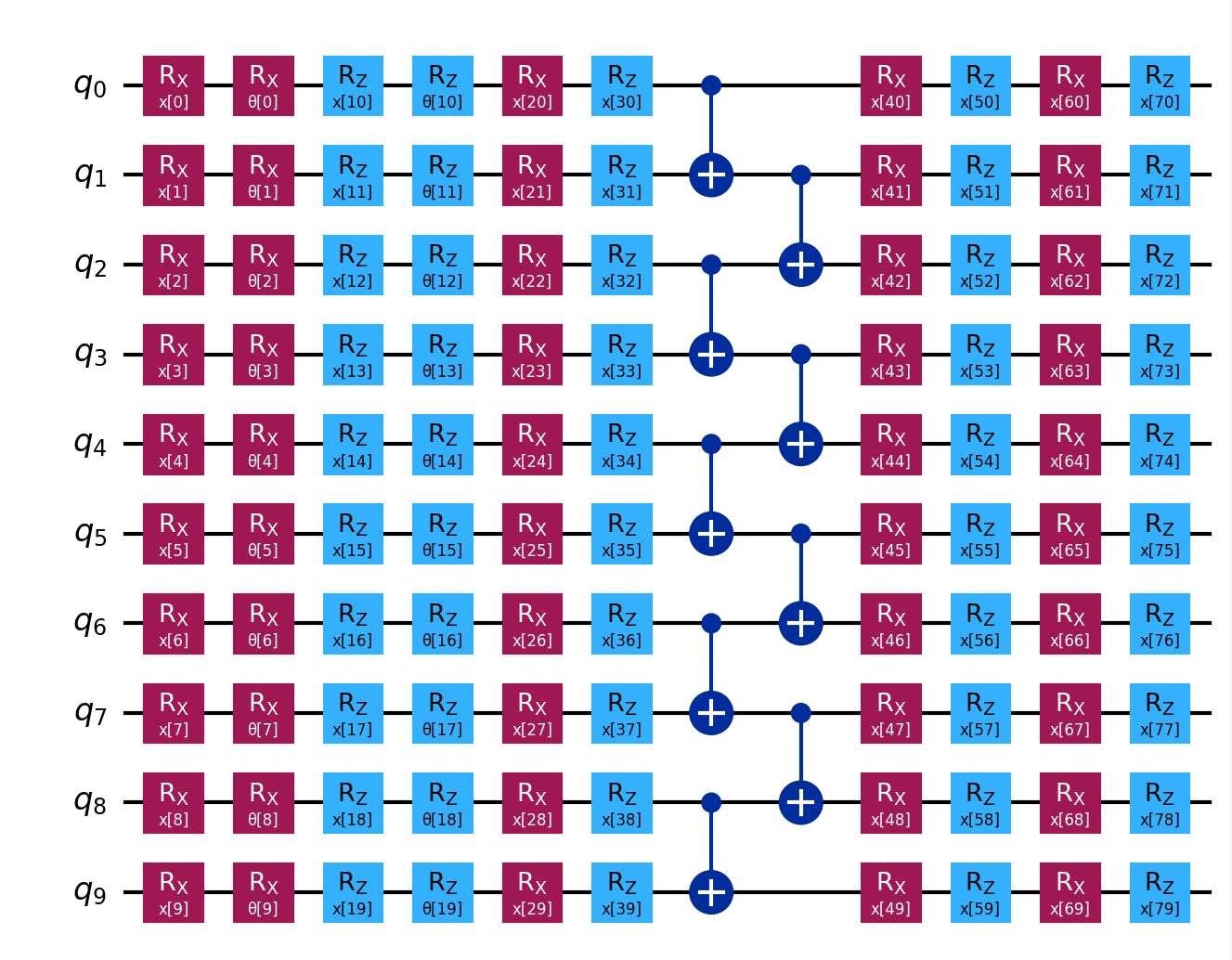}
	\caption{
		Compact feature map for compressed input data, reduced via PCA
		transformation. The embedding circuit consists of 80 angle encoding gates,
		loading the classical vector $(x[0],\dots, x[{79}])$ into the quantum register, 
		and 20 parametric gates, with parameters $(\theta[0],\dots, \theta[{19}])$. Note that the parametric 
		gates can be combined with the encoding gates to reduce the circuit depth, and further merged
		via the Euler angle decomposition. 
	}
	\label{fig:featurecompat}
\end{figure}

However, such a deep quantum circuit 
is challenging in terms of both implementability with NISQ hardware
and trainability of the parameters. 
For these reasons, we decided to work also with a more compact circuit, shown in Fig.~\ref{fig:featurecompat},
that we call \textit{compact feature map}. 
By means of a Principal Component Analysis (PCA) carried on
the training data we identified that with a compression of the data space from
a dimension of 256 to 80 it is still possible to convey more than 94\% of the
information, as measured by the cumulative sum of the variance explained by the selected components. We also introduce 20 tunable angles $\theta_i$ with $i=0\ldots 19$ and encode
them together with the input data $x_i$, in the first two layers of the quantum
circuit. This choice should reduce the expressibility of the associated feature
map, which is a necessary condition to avoid concentration of quantum
kernel values phenomenon \citep{thanasilp2022exponential}, an
obstacle to trainability of kernel methods.


\begin{figure}[t]
	\centering
	\includegraphics[width=0.4\textwidth]{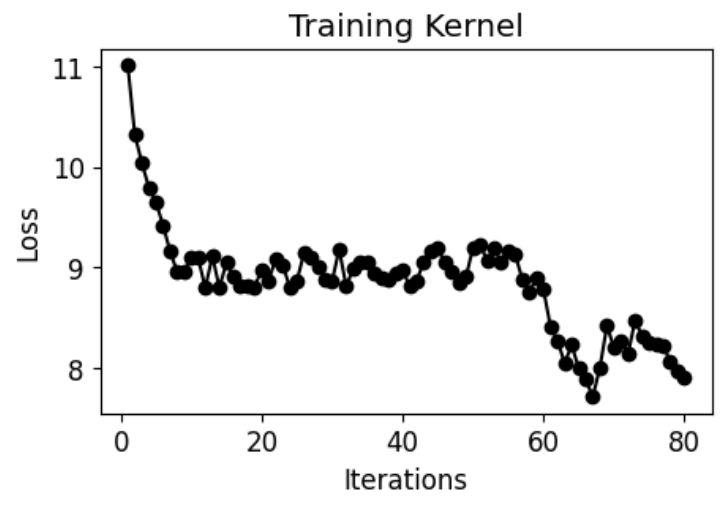}
	\includegraphics[width=0.4\textwidth]{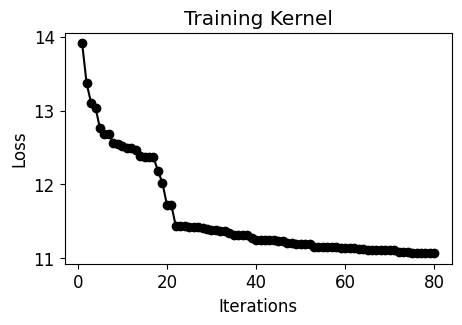}
	\caption{Kernel-target alignment of both {large} (top figure) and {compact} 
	(bottom figure) feature maps. The Cross-entropy loss was employed for the min-max optimization.}
	\label{fig:alignment}
\end{figure}

\subsection{Standard training} 

We carried out the kernel alignment procedure, described in appendix \ref{sec:kernel alignement}, to 
optimize the tunable parameters of the quantum feature maps. The decay of the cross-entropy loss 
as the number of iterations increases is shown in Fig.~\ref{fig:alignment}, 
for both the large and compact quantum embeddings. 
The optimization of the parameters was quite computationally heavy. We
performed the kernel alignment with respect to the full training dataset, which
required the evaluation of the quantum circuit many times to compute a 500 by
500 kernel matrix needed in each iteration step of the optimization procedure.
Both circuits have been initialized with random parameters and optimized with 
SPSA. More details are provided in appendix~\ref{sec:alignment
numerics}.

\begin{figure}[t]
	\centering
	\includegraphics[height=0.15\textheight]{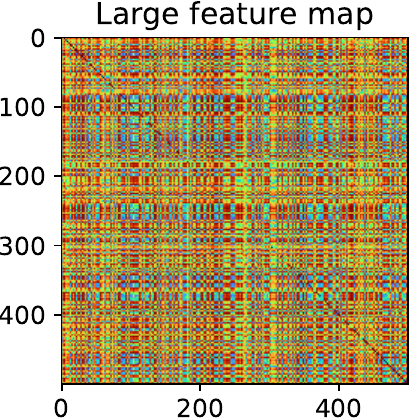}
	~~~
	\includegraphics[height=0.15\textheight]{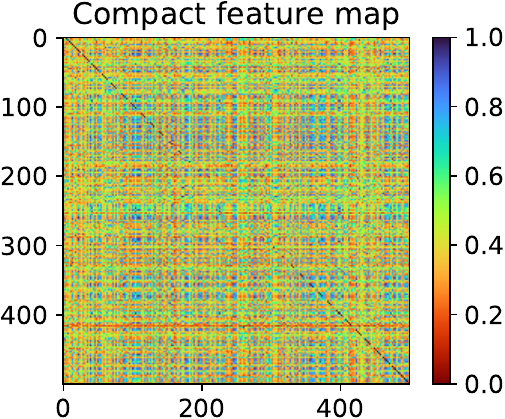}
	\caption{Kernel matrices of the training dataset, after the kernel alignment procedure,
		using either the large (left figure) or the compact (right figure) feature maps.
	}
	\label{fig:kernelmatrix}
\end{figure}

Once the parameters of the quantum embedding circuits are optimized with the
kernel alignment procedure, we can observe that the entries of the kernel matrices, shown 
in Fig.~\ref{fig:kernelmatrix}, 
are quite distributed within the interval $[0,1]$ for both quantum embeddings
circuits. This means that we are far away from a situation where we can
encounter the exponential concentration phenomenon, where 
we would expect to find large areas of the kernel matrix concentrated
around a fix value, with an associated image that takes the same color in large
continuous zones above and below the diagonal.

The kernel matrices computed by means of the larger feature map contain several
``spots'' where the kernel value is zero. This is probably because the kernel
alignment procedure was interrupted after 80 steps, when the loss function was
still decreasing. This is a symptom that the kernel alignment procedure was not
completed and that there is still margin to optimize the parameters of that
quantum embedding circuit.

\begin{table}[t]
	\begin{tabular}{|c|c|c|} \hline
 			&Large feature map  & Compact feature map \\ \hline 
		Training accuracy & 1.00 & 1.00\\\hline 
		Test accuracy & 0.99 & 1.00\\\hline 
		F1-score (test) & 0.99 & 1.00\\\hline 
	\end{tabular} 
	\caption{Summary of the performance of trained QSVM models.}
	\label{tab:trained svm}
\end{table}

\begin{table}[t]
	\begin{tabular}{ c|c|wc{15mm}|wc{15mm}|wc{15mm}|wc{15mm}| } 
		\multicolumn{2}{c}{} & \multicolumn{2}{c}{Large feature map} & 
		\multicolumn{2}{c}{Compact feature map} \\ \cline{2-6}
		\multirow{4}{*}{\rotatebox{90}{True}} & & Hand & Breast & Hand & Breast  \\ \cline{2-6}
		& Hand\phantom{$^{3^3}$}& \cellcolor{blue!25}{45} & 0&\cellcolor{blue!25}{45} & 0  \\ \cline{2-6}
		& Breast\phantom{$^{3^3}$} & 1 & \cellcolor{blue!25}{54} & 0 & \cellcolor{blue!25}{55} \\ \cline{2-6}
	\end{tabular}
	\caption{ Confusion matrix between true (T) and predicted (P) labels, 
		evaluated on the test dataset for the trained QSVM based on the 
		large (left) and compact (right) feature maps.
	}
	\label{tab:confusion}
\end{table}


After computing the $500\times500$ kernel matrix, 
we can fit a classical SVM algorithm on the
training dataset, exploiting the kernel computed via a quantum circuit. 
The performance of the trained SVM models is shown in Table~\ref{tab:trained svm},
while the confusion matrix is illustrated in Table.~\ref{tab:confusion}.
Both SVM classifiers have been trained adopting a K-fold cross-validation strategy with $K=5$ 
to select the hyper-parameter $C$, with found optimal value $C=1$.

\subsection{Adversarial training}

\begin{figure*}[t]
	\centering
    \includegraphics[width=1.0\textwidth]{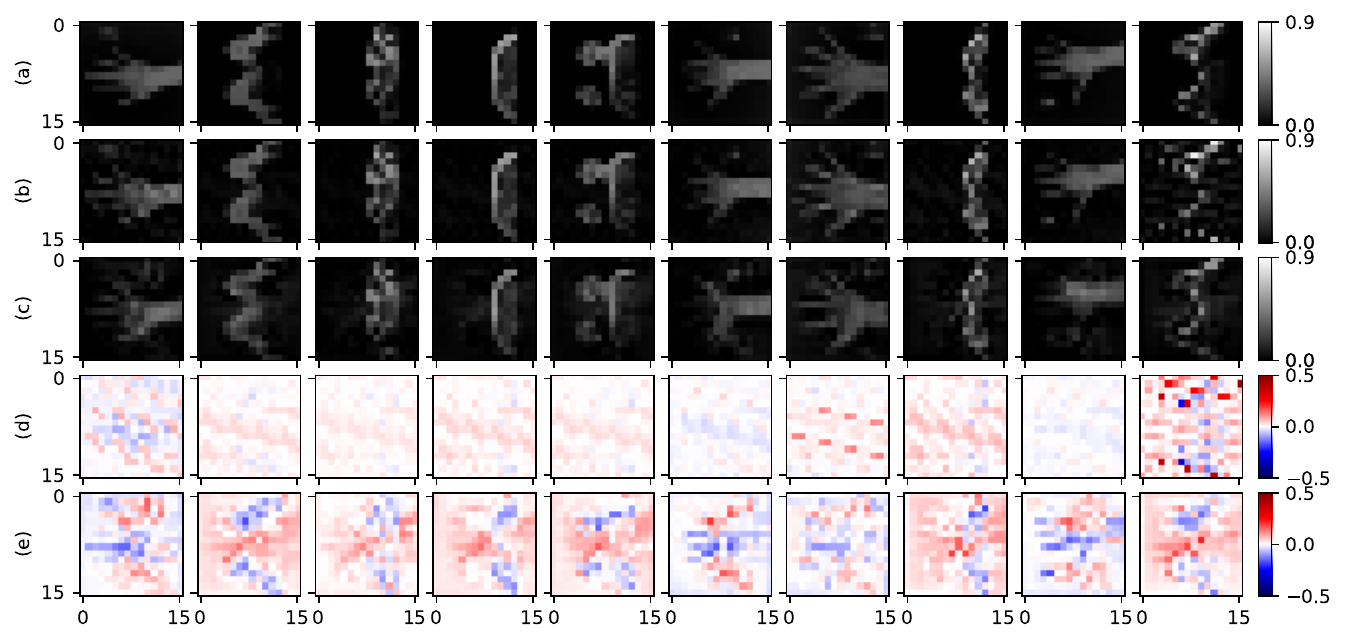}
	\caption{Original and adversarial images arranged into a grid. In each column, (a) is the original sample, (b) and (c) the corresponding adversarial examples for the large and compact feature map, respectively, (d) and (e) show the perturbation that, summed up to the original image, generates (b) and (c), respectively. 
	}
	\label{fig:adv images}
\end{figure*}

After training the QSVM, we produce adversarial images by following the procedure outlined in Sec.~\ref{sec:qadv}. Some examples are shown in Fig.~\ref{fig:adv images}. All the produced adversarial examples deceive the QSVM classifier by construction. In most cases, a human eye can still recognise an adversarial hand as a hand, and an adversarial breast as a breast. However, in a few examples, e.g. the sixth, ninth and tenth columns in Fig.~\ref{fig:adv images}, the algorithm smoothly transform a hand into a breast, signalling that the algorithm introduced a large perturbation to the image.

We generate 50 adversarial images for both large and compact embeddings and
randomly distribute them using a stratified approach into two sets, one with 41
samples 
and one with the remaining 9 samples. The larger set is then joined to the original training samples, while the smaller one to the test samples.

\begin{table*}[t]
	\begin{tabular}{|c|c|c|}
		\multicolumn{3}{c}{} \\ \hline
 			&Large feature map  & Compact feature map \\ \hline 
		Test accuracy & 0.91 & 0.92\\\hline 
		F1-score (test) & 0.92 & 0.92\\\hline 
	\end{tabular} 
	~~~~
	\begin{tabular}{ c|c|wc{15mm}|wc{15mm}|wc{15mm}|wc{15mm}| } 
		\multicolumn{2}{c}{} & \multicolumn{2}{c}{Large feature map} & 
		\multicolumn{2}{c}{Compact feature map} \\ \cline{2-6}
		\multirow{4}{*}{\rotatebox{90}{True}} & & Hand & Breast & Hand & Breast  \\ \cline{2-6}
		& Hand\phantom{$^{3^3}$}& \cellcolor{blue!25}{45} & 4&\cellcolor{blue!25}{45} & 4  \\ \cline{2-6}
		& Breast\phantom{$^{3^3}$} & 6 & \cellcolor{blue!25}{54} & 5 & \cellcolor{blue!25}{55} \\ \cline{2-6}
	\end{tabular}
	\caption{Summary of the performance of trained QSVM models on the test set extended with adversarial examples,
		together with the confusion matrix between true and predicted labels, 
		evaluated on the test dataset extended with 9 adversarial samples (4 'hand' and 5 'breast') for the trained QSVM based on the 
		large and compact feature maps.
	}
	\label{tab: confusion qsvm deceived}
\end{table*}

As expected, checking the behaviour of the trained QSVM classifier on the extended test dataset reveals a decrease of the performance. As detailed in Table.~\ref{tab: confusion qsvm deceived} all the perturbed examples are wrongly classified. This highlights the vulnerability of the previously trained QSVM to evasion attacks.



The next step is to verify whether the adversarial training technique results effective for the development of QSVM classifiers robust to evasion attacks, in analogy to what was previously observed for the QNN classifier \citep{ren2022experimental}.

We carry on a new training of the QSVM considering the modified training dataset, now extended with 41 adversarial examples. 
We keep the parameters of the quantum embedding fixed to the values we have
optimized before, via the quantum alignment. Also in this case we adopted a
K-fold cross-validation strategy with K=5 to identify the best hyper-parameter
C. The result is C=1.0 for the large embedding and C=2.0 for the compact one.
After adversarial training with both models, 
we observe a significant increase of the performance, as detailed in 
Table.~\ref{tab:confusion adv}.

\begin{table*}[t]
	\begin{tabular}{|c|c|c|} \hline
 			&Large feature map  & Compact feature map \\ \hline 
		Training accuracy & 0.99 & 0.99\\\hline 
		Test accuracy & 1.00 & 1.00\\\hline 
		F1-score (test) & 1.00 & 1.00\\\hline 
	\end{tabular} 
	~~
	\begin{tabular}{ c|c|wc{15mm}|wc{15mm}|wc{15mm}|wc{15mm}| } 
		\multicolumn{2}{c}{} & \multicolumn{2}{c}{Large feature map} & 
		\multicolumn{2}{c}{Compact feature map} \\ \cline{2-6}
		\multirow{4}{*}{\rotatebox{90}{True}} & & Hand & Breast & Hand & Breast  \\ \cline{2-6}
		& Hand\phantom{$^{3^3}$}& \cellcolor{blue!25}{49} & 0&\cellcolor{blue!25}{49} & 0  \\ \cline{2-6}
		& Breast\phantom{$^{3^3}$} & 0 & \cellcolor{blue!25}{60} & 0 & \cellcolor{blue!25}{60} \\ \cline{2-6}
	\end{tabular}
	\caption{
	Summary of the performance of QSVM models after adversarial training, together with the 
	confusion matrix between true and predicted labels, evaluated on the extended test dataset after adversarial training.}
	\label{tab:confusion adv}
\end{table*}

Looking at these numerical results we can affirm that the adversarial training technique is effective for the development of QSVM classifiers robust to evasion attacks.

\subsection{Kernel computation on a real quantum system}
\label{sec: real quantum system}

In this section we describe a proof-of-principle experiment on 
a real quantum hardware. We consider the adversarially trained QSVM model based on the
compact embedding, and use a real quantum hardware to
estimate the quantum kernel values required to predict the labels of previously
unseen samples. For this experiment we used a quantum computer 
available within the open plan service of the 
\textit{IBM Quantum Platform}. 
Because of the limitations we had on the resources available, we
leveraged the quantum hardware to predict the label of one sample only.
Nevertheless, the procedure is general and valid to perform computations for
any number of samples.




To reduce computation time on the quantum system and mitigate issues 
associated with noise, we transpiled the circuit
considering the specificities of the real hardware. Among the quantum systems
accessible via the \textit{IBM Quantum Platform}, we selected the \verb+ibm_kyoto+
backend, a quantum system based on the quantum processor family \verb+Eagle+,
with 127 qubits and a quantum volume of 128.
The final circuit is characterized by 77 layers of gates.


To estimate the quantum kernel values on the real quantum hardware we used the \verb+Sampler+ primitive 
from Qiskit \citep{Qiskit}, configured to carry on 1024 shots for each circuit. 
The quantum processor evaluated 553984 ($1024\times541$) circuits and took 8 minutes
approximately. Once the job was completed and the results provided back, we
populated the kernel matrix.

Finally, we used the nonlinear SVM previously trained to predict the label of
the sample. Our experiment terminated successfully: the adversarial example was
classified correctly by the quantum-enhanced SVM, trained adopting the
adversarial training technique and evaluating the quantum kernel on a real
quantum processor.

This proof-of-principle experiments showcase the possibility that adversarial training 
makes the predictions of a QSVM model more robust, not only against perturbations of the inputs,
but also against hardware noise \citep{banchi2022robust}.

\section{Conclusions}

The numerical results from our simulations clearly show that quantum
classifiers based on kernel methods, like quantum-enhanced SVM, are vulnerable
to the evasion attacks, similarly to what happens with Quantum Neural Networks
\citep{ren2022experimental}. 
The well-developed mathematical theory behind SVM allows us to have 
a simple geometrical picture to understand when this is possible. 
The narrower the SVM margin is, the smaller is the perturbation that an attacker needs to generate,
and so the higher is the possibility to disguise an adversarial example as a legitimate sample.
Due to the kernel concentration phenomenon \citep{thanasilp2022exponential}, this possibility is quite 
likely with complex quantum embeddings. 

We have also shown how to make the prediction robust against such attacks by
using adversarial training:
adding adversarial examples to the training dataset
makes the SVM algorithm to select an optimal decision hypersurface with better
generalization and, hence, better distinguishing capabilities. In 
other terms, adversarial
training forces  the decision hypersurface to position itself in the Hilbert
space in such a way to achieve the largest margin.
While from the kernel concentration phenomenon we may expect, for complex
embeddings, a high success probability in carrying out evasion attacks, 
the robustness of adversarial training in such regime is less clear and 
needs further studies. Indeed, 
the concentration depends also on the
expressibility induced by the input dataset. 
On one hand we would like to
enlarge the legitimate training dataset with as many adversarial examples as
possible, so to improve the robustness of the classifier, but on the other hand 
such an increase induces an increase in the input randomness, which 
might make kernel values even more concentrated and, as such, more difficult 
to evaluate experimentally. 
From a general perspective, this  can be considered as a
consequence of the fact that it becomes extremely difficult to extract any
useful information from the large Hilbert space. 

Finally, we have conducted a proof-of-principle experiment on a real quantum hardware, 
which suggests that adversarial training can be a useful tool to make quantum kernel 
methods robust, not just against input perturbations, but also against hardware noise.

As another future perspective, it would be interesting to link the generalization capabilities 
of quantum kernel methods and their adversarial robustness. 
The generalization capabilities of the quantum-enhanced SVM classifier are
defined intrinsically by the quantum embedding circuit leveraged for quantum
kernel estimation, and can be bounded in several ways 
\citep{banchi2021generalization,banchi2023statistical,huang2021power,caro2022generalization}. 
Moreover, a link between generalization and adversarial learning was recently considered 
in \cite{georgiou2024adversarial}. 
However, 
there are currently no empirical guidelines 
to select an appropriate quantum embedding circuit for a specific classification task.



\section*{Acknowledgements}

L.B.~acknowledges support from:
Prin 2022 - DD N.~104 del 2/2/2022, entitled ``understanding the LEarning process of QUantum
Neural networks (LeQun)'', proposal code 2022WHZ5XH, CUP B53D23009530006;
U.S. Department of Energy, Office of Science, National Quantum Information Science Research
Centers, Superconducting Quantum Materials and Systems Center (SQMS) under the Contract
No. DE-AC02-07CH11359;
PNRR Ministero Università e Ricerca Project No. PE0000023-NQSTI.

\appendix

\section{Quantum Kernel Alignment}\label{sec:kernel alignement}

Aside from some specific cases, for instance where datasets to consider belong
to a group \citep{glick2024covariant}, the selection of a kernel
and the associated feature map is not immediate, and no general guideline is available.
In a context like ours, where the quantum embedding is a parameterized circuit,
looking for a ``good'' kernel means looking for a parameters set that enables to
obtain a ``good'' classifier. 
Common metrics to evaluate the quality of a classifier, like
accuracy, precision, recall, etc., do not fit well for this purpose because
these are discrete metrics that do not allow to detect small improvements due
to a parameters change, for example in terms of generalization, and that
approach would require to perform an exhaustive search in the parameter space
without any hint on regions where to search. A specialized measure that we can
use is the \emph{kernel-target alignment}. Such a measure derives directly
from the \emph{kernel alignment} \citep{cristianini2001kernel,wang2015overview}
that is defined as:
\begin{equation}
A\left(\mathrm{K}_{1},\mathrm{K}_{2}\right)=\frac{\left\langle \mathrm{K}_{1},\mathrm{K}_{2}\right\rangle }{\sqrt{\langle \mathrm{K}_{1},\mathrm{K}_{1}\rangle \langle \mathrm{K}_{2},\mathrm{K}_{2}\rangle }},
\end{equation}
where $\mathrm{K}_{1}$ and $\mathrm{K}_{2}$ denote the kernel matrix (Gram matrix) of two kernels $k_{1}\left(\cdot ,\cdot \right)$ and $k_{2}\left(\cdot ,\cdot \right)$ defined over a certain dataset, and $\langle \mathrm{K}_{1},\mathrm{K}_{2}\rangle $ is the Frobenius inner product between two matrices, which is given by
\begin{equation}
\left\langle \mathrm{K}_{1},\mathrm{K}_{2}\right\rangle =\sum _{i,j}k_{1}\left({x}_{i},{x}_{j}\right)k_{2}\left({x}_{i},{x}_{j}\right)
\end{equation}
Kernel alignment measures the similarity between two kernels and we can give it a geometrical interpretation: considering $\mathrm{K}_{1}$ and $\mathrm{K}_{2}$ as two vectors in a bi-dimensional vector space, the kernel alignment is the cosine of the angle between the two vectors. Given that is possible to show that a kernel matrix is a positive semidefinite matrix, the kernel alignment takes values between 0 and 1, with 1 indicating maximal similarity.

The kernel-target alignment measure is defined as the kernel alignment between a kernel matrix of our interest and an ideal target kernel matrix defined as $\mathrm{K}^{\mathrm{\imath }}=\mathrm{y}\mathrm{y}^{T}$, which represents a kernel function that assigns to each couple of samples the product of the corresponding labels: $k^{\mathrm{\imath }}\left({x}_{i},{x}_{j}\right)=y_{i}y_{j}$. It can be described as:
\begin{equation}
    A\left(\mathrm{K},\mathrm{K}^{\mathrm{\imath }}\right)=\frac{\left\langle \mathrm{K},\mathrm{y}\mathrm{y}^{t}\right\rangle }{\sqrt{\langle \mathrm{K},\mathrm{K}\rangle \langle \mathrm{y}\mathrm{y}^{t},\mathrm{y}\mathrm{y}^{t}\rangle }}=\frac{\mathrm{y}^{t}\mathrm{K}\:\text{y}}{M\| \mathrm{K}\| }
    \label{eq: kernel target alignment}
\end{equation}
where $M$ is the number of elements in the labels vector $\mathrm{y}$, that is the number of samples in the dataset.

The kernel-target alignment measures how well a kernel function reproduces the behaviour of the ideal target kernel, and is characterized by some properties:
\begin{enumerate}
	\item It is computationally efficient compared to other metrics \citep{chapelle2002choosing}.
		It can be computed in $O\left(M^{2}\right)$ time complexity.
\item It enjoys the concentration property, which means that the kernel-target alignment is concentrated around its mean value. Because of this, an empirical estimate of it is close to the true alignment value.
\item It provides theoretical basis to the generalization capabilities of a binary classifier that uses a kernel with an high mean value of the kernel-target alignment. Optimizing this alignment provides an upper bound to the generalization error (the probability a trained model will predict wrong labels for unseen test samples) of a simple classifier based on center of mass of the data classes.
\end{enumerate}
To tune the quantum embedding parameters and align the associated kernel to the ideal target one we need to solve an optimization problem: find the optimal parameters that maximize the kernel-target alignment
\begin{equation}
\theta ^{\mathrm{\imath }}=\argmax_\theta A\left(\mathrm{K}_{\theta },\mathrm{K}^{\mathrm{\imath }}\right),
\end{equation}
where $\mathrm{K}_{\theta }$ is the Gram matrix for the parameterized kernel $k_{\theta }\left(\cdot ,\cdot \right)$.

Assuming that the kernel is differentiable, it is possible to show that kernel-target alignment is differentiable with respect to the parameters vector and therefore it is possible to use a gradient based algorithm to optimize the parameters. Clearly, since we want to maximize the alignment, we have to minimize it’s opposite, that means the kernel-target alignment multiplied by -1, or use a gradient ascent algorithm.

For this project we decided to exploit a slightly different method to implement the kernel alignment. In particular we followed the approach described in \cite{glick2024covariant}. Instead of maximizing the kernel-target alignment, we look for the parameters set that minimize the upper boundary for the generalization error. In particular, the object function of the optimization problem to consider, coincides with the decision function of the nonlinear SVM algorithm, in the dual problem formulation:
\begin{equation}
    F\left(\alpha ,\theta \right)=\sum _{i}\alpha _{i}-\frac{1}{2}\sum _{i,j}\alpha _{i}\alpha _{j}y_{i}y_{j}k_{\theta }\left(\mathrm{x}_{i},\mathrm{x}_{j}\right)
    \label{eq: kernel align objfunc}
\end{equation}
Fitting a SVM model entails the maximization of this function under constraints, as shown by
Eq.~\eqref{eq: dual opt}, while optimizing for generalization requires
minimizing this same function over the kernel parameters $\theta $. This
min-max problem can be interpreted as choosing a kernel among all the possible
$\mathrm{K}_{\theta }$ that minimize the SVM bound on the classification error.
In terms of complexity, this metrics is less efficient than kernel-target
alignment, since in addition to the evaluation of a Gram matrix in
$O\left(M^{2}\right)$ we have also to solve the dual optimization problem for
maximizing the margin of a SVM model. Nevertheless, this approach allows to
reduce the boundary for the generalization error of an SVM classifier, and this
is an important aspect for the purposes of this work.

\section{Details of numerical results} \label{sec:alignment numerics}

Regarding the experiments on the kernel alignment procedure we can report some observations:
\begin{enumerate}
    \item The objective function $F\left(\alpha ,\theta \right)$ ~\eqref{eq: kernel align objfunc} can be seen as a function enclosing a weighted version of ~\eqref{eq: dual opt} (with $\alpha_{i}\alpha_{j}$ as weights) and it is differentiable respect to both $\alpha $ and $\theta$. We did some simulations to compare the use of SPSA and COBYLA as optimizers. The use of a gradient-based optimization algorithm like SPSA looks generating smaller fluctuation of the objective function compared to the use of COBYLA. SPSA is indeed proposed as default optimizer for a kernel alignment procedure based on the qiskit’s \textit{QuantumKernelTrainer} tool.
    \item The optimization process was carried on by means of SPSA optimizer for 80 iterations, with learning rate=0.05 and perturbation=0.05 on both feature maps. While the loss of the compact quantum embedding decreases quite smoothly and stabilizes around a value of 11, the large embedding shows a rougher evolution reaching a final value of approximately 8.
    \item Considering that we executed the kernel alignment on a quantum simulator without noise, we could have used a vanilla gradient descent optimizer instead of SPSA. However, with the perspective to use this same method on a real quantum hardware affected by noise, the use of a stochastic optimization algorithm is expected to be necessary to reduce non convergence issues.
    \item It’s curious to observe that the property of ``concentration'' described in \citep{cristianini2001kernel} in a positive sense and exploitable to obtain kernels that allow SVM-based classifiers to generalize well, it could be some how connected to the ``exponential concentration'' phenomenon that affects quantum kernel methods in a negative sense and that makes difficult to carry on the training of parameterized quantum kernels \citep{thanasilp2022exponential}. 
\end{enumerate}

Regarding the experiments on adversarial learning:
\begin{enumerate}
	\item We generated a set of 50 adversarial examples for both QSVM classifiers based on large and compact embedding, respectively, by means of the iterative process described by ~\eqref{eq: gradient iteration}, with $\eta=0.01$.
    \item The algorithm takes less then 50 iterations to generate an adversarial example for the compact embedding while, for the large embedding, it takes approximately 1270 iterations in the worst case.
 
\end{enumerate}

\section{Alternative technique to generate adversarial samples}\label{sec:old adv samples}

To deal with a general quantum kernel we have to identify a different technique that does not require the knowledge about the gradient of the kernel. A possible approach consists on setting up an optimization problem where we use an objective function that once minimized satisfies the same goals of the adversarial learning technique for QSVM. To generate a good adversarial sample starting from an unperturbed input sample we have to satisfy two conditions simultaneously (as already mentioned in section~\ref{sec:qadv}):
\begin{enumerate}
    \item
    \label{item: first goal}
    To perturb an input sample $\mathrm{x}$ into its adversarial version $\mathrm{x}_{adv}$ so that its associated vector $\rho \left(\mathrm{x}\right)$ in the larger feature space gets shifted to $\rho \left(\mathrm{x}_{adv}\right)$ on the opposite side of the optimal decision plane defined by the QSVM classifier.
    \item
    \label{item: second goal}
    To keep the distance between $\mathrm{x}$ and $\mathrm{x}_{adv}$ as small as possible.
\end{enumerate}
To foster the achievement of the first goal we can define a function satisfying that same condition when it gets minimized. There are clearly infinite ways to do that. One possibility is to define a function like:
\begin{align}
    \label{eq: objfunc g1}
	g_{1}\left(\mathrm{x}_{adv}\right)&=\max\left(0,1-y_{i}r_{adv}\right)=\\&=\max\left(0,1-y_{i}\frac{\left\langle \mathrm{w},\rho \left(\mathrm{x}_{adv}\right)\right\rangle +b}{\sqrt{\langle \mathrm{w},\mathrm{w}\rangle }}\right)=\\&=\max\left(0,1-y_{i}\frac{f\left(\mathrm{x}_{adv}\right)}{\parallel \mathrm{w}\parallel ^{2}}\right),
\end{align}
where $r_{adv}$ is the distance of the feature vector $\rho \left(\mathrm{x}_{adv}\right)$ from the nonlinear SVM's decision plane, defined by the tuple $(w, b)$, and $f\left(\mathrm{x}_{adv}\right)$ is the decision function of the SVM.

The nonlinear function ~\eqref{eq: objfunc g1} takes values zero when $\rho \left(\mathrm{x}_{adv}\right)$ is, simultaneously, at a distance greater than or equal to 1 from the decision plane $(w, b)$ and on the opposite side with respect to $\rho \left(\mathrm{x}\right)$. Otherwise, the function takes a positive value which magnitude increases linearly while $\rho \left(\mathrm{x_{adv}}\right)$ moves toward the the side of the decision plane where $\rho \left(\mathrm{x}\right)$ lies. Minimizing ~\eqref{eq: objfunc g1} implies searching for an $\mathrm{x}_{adv}$ for which the SVM classifier predicts a label different from the unperturbed sample's one.

\begin{figure*}[ht!]
	\centering
    \includegraphics[width=1.0\textwidth]{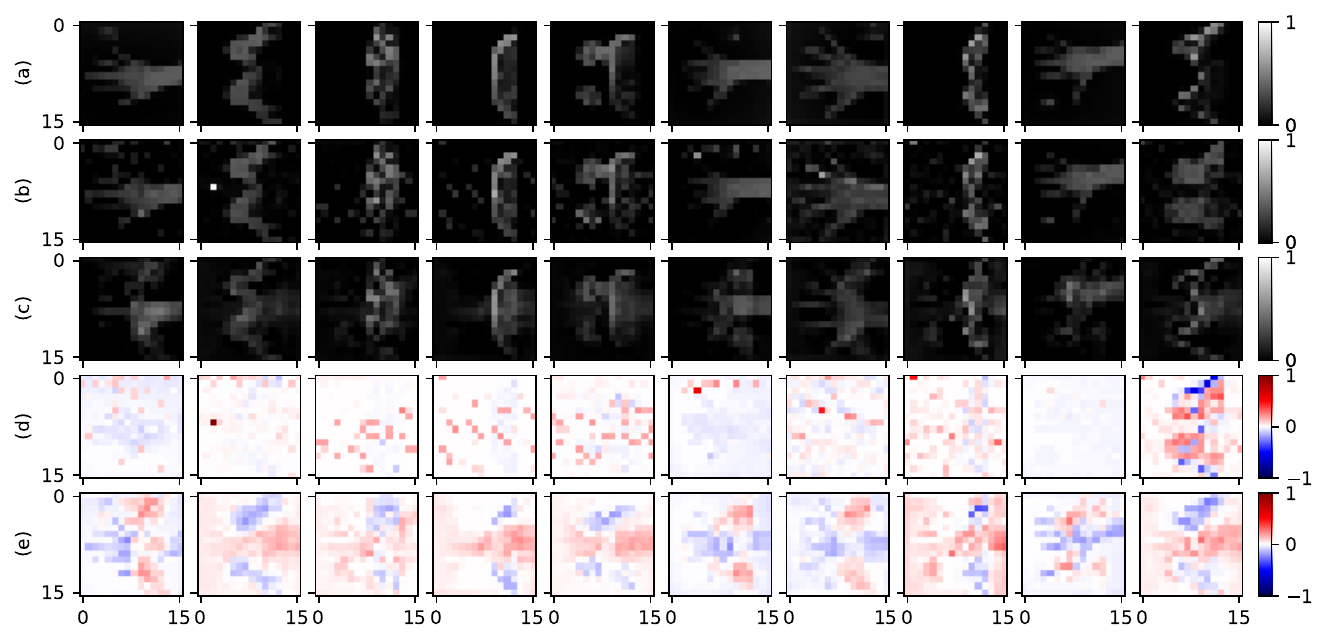}
	\caption{Example of adversarial images generated with the alternative adversarial learning technique, with $\epsilon =3.0$ and $0.5 \leq \mu \leq 2.0$. In each column, (a) is the original sample, (b) and (c) the corresponding adversarial examples for the large and compact feature map, respectively, (d) and (e) the perturbation that summed up to the original image generates (b) and (c), respectively.}
	\label{fig: alternative adv images}
\end{figure*}

To pursue the second goal we can define a function that has a minimum when the distance ${\rm dist}(\mathrm{x},\mathrm{x}_{adv})$ is minimized. For instance we can consider the squared distance:
\begin{align}
    \label{eq: objfunc g2}
    g_{2}\left(\mathrm{x}_{adv}\right)=\parallel \mathrm{x}-\mathrm{x}_{adv}{\parallel }_{2}^{2}
\end{align}
We can define an overall objective function for our optimization problem as a combination of the previous two functions, where we introduce some parameters that we can use to tune empirically the minimization process:
\begin{align}
    \label{eq: objfunc g}
		&g\left(\mathrm{x}_{adv},\epsilon ,\mu \right)=g_{1}\left(\mathrm{x}_{adv}\right)^{\epsilon }+\mu g_{2}\left(\mathrm{x}_{adv}\right)= \\&~~~~=\left(\max\left(0,1-y_{i}\frac{f\left(\mathrm{x}_{adv}\right)}{\parallel \mathrm{w}\parallel }\right)\right)^{\epsilon }+\mu \parallel \mathrm{x}-\mathrm{x}_{adv}{\parallel }_{2}^{2}\nonumber
\end{align}
The legitimate training sample $\mathrm{x}$ is considered a constant.
The objective function $g\left(\mathrm{x},\epsilon ,\mu \right)$ is a nonlinear function with respect to $\mathrm{x}_{adv}$ and therefore we can only aim to find local optimal values. Furthermore, considering that the QSVM's decision function $f\left(\mathrm{x}_{adv}\right)$ is nonlinear, nothing prevents function ~\eqref{eq: objfunc g} to have one or more local/global minimums where ~\eqref{eq: objfunc g1} is different from zero.

This technique to generate adversarial samples is therefore an empirical process where parameters $\epsilon$ and $\mu$ need to be tuned in a trial and error iterative process that brings to generate adversarial examples while keeping ~\eqref{eq: objfunc g2} limited.
Some adversarially generated samples are shown in Fig.~\ref{fig: alternative adv images}.

\end{document}